\documentclass{ws-ijmpa}
\newcommand{\sls}[1]{#1\!\!\!/}

\begin{document}
\markboth{Shou-Shan Bao}{The Mass Difference of $F-\bar{F}$ with $SO(3)$ Family Gauge Symmetry}

%
\catchline{}{}{}{}{}
%

\title{The Mass Difference of $F-\bar{F}$ with $SO(3)$ Family Gauge Symmetry}
\author{Hong-Lei Li}
\address{School of Physics and Technology, University of Jinan, Jinan Shandong 250022, P.R.China\\and\\School of Physics, Shandong University, Jinan Shandong 250100, P.R.China\\ lihl@mail.sdu.edu.cn}

\author{Shou-Shan Bao}
\address{School of Physics, Shandong University, Jinan Shandong 250100, P.R.China\\ ssbao@sdu.edu.cn}
\author{Zong-Guo Si}
\address{School of Physics, Shandong University, Jinan Shandong 250100, P.R.China\\ zgsi@sdu.edu.cn}
\maketitle
\begin{history}
\received{Day Month Year}
\revised{Day Month Year}
\end{history}
\begin{abstract}
 As a simple extension, a non-Abelian family gauge symmetry SO(3), as well as three family Majorana neutrinos, was introduced to explain the tri-bimaximal mixing matrix of neutrinos.  We discuss the effect of the possible $SO(3)$ family gauge interaction to the mass differences of $K-\bar{K}$, $B_d-\bar{B}_d$, $B_s-\bar{B}_s$ and $D-\bar{D}$, and get the constrains to the new gauge bosons.
\end{abstract}
\ccode{12.60.Cn; 11.30.Hv; 14.40.Lb; 14.70.Pw}

\section{Introduction}
Though the Standard Model (SM) has obtained great success in explaining the phenomena of particle physics, the mass spectra of quarks and leptons still remains a profound mystery. The existence of massive neutrinos is a challenge to SM. To understand the mass spectra and mixing of quarks and leptons, three family Majorana neutrinos and family gauge symmetry $SO(3)_f$ have been introduced as a simple extension to SM\cite{Sakharov:1994pr,Wu:1998if,Wu:1999dh,Wu:1999mu}. In this paper we focus on the $\Delta F=2$ processes with such new gauge interaction.

The current neutrino experiments \cite{EXP1,EXP2,EXP3,EXP4,EXP5,EXP6,EXP7,EXP8,SV} can be well described by
neutrino oscillation via the mixing of three neutrinos\cite{SV,TNM2,TNM1}. The global fit leads to tiny neutrino masses
and large mixing\cite{SV}. Phenomenologically, such mixing angles are consistent with the so-called tri-bimaximal
mixing with $\theta_{12}=35^\circ$, $\theta_{23}=45^\circ$ and $\theta_{13}=0$, which was first proposed by Harrison,
Perkins, and Scott\cite{HPS}. As the mixing matrix is symmetric, many theoretical efforts have been made to
obtain such a mixing matrix via imposing various symmetries, especially the discrete symmetries,
which in general lead $\theta_{13}=0$. Recently, it was shown that the tri-bimaximal matrix may be regarded as the lowest order approximation of the neutrino mixing matrix, where the mixing angle $\theta_{13}$ is nonzero and can be tested experimentally\cite{Abe:2011sj,An:2012eh}. With three family neutrinos, the gauge family symmetry $SO(3)$ instead of discrete symmetries was discussed
in\cite{Wu:1999yz,Wu:2008zzj,Wu:1999yg,CS,Ma:1998db,Wetterich:1998vh,Barbieri:1999km}. In this case, the tri-bimaximal neutrino mixing matrix can be obtained as the lowest order approximation from diagonalizing a symmetric mass matrix\cite{Wu:2008zzj}.

As the Higgs boson has not been confirmed yet, the electro-weak (EW) symmetry breaking mechanism is unclear. One can suppose the family gauge symmetry spontaneously violates with the EW symmetry, which implies different Yukawa terms from those in SM as shown in\cite{Sakharov:1994pr,Wu:1998if,Wu:1999dh,Wu:1999mu,Berezhiani:1990wn,Berezhiani:1990jj,Berezhiani:1990sy,Berezhiani:1989fs,Berezhiani:1989fp,Sakharov:1996un}. We work with the low energy effective Lagrangian after the gauge symmetries breaking, and discuss the effect of the new gauge interaction to the mass differences of the neutral mesons. When the family gauge symmetry is introduced, there will be flavor changing neutral current (FCNC) at tree level, especially the $\Delta F=2$ processes. The mass difference has been observed in the neutral pseudo-scalar meson systems, denoted as $F-\bar{F}$, where $F$ refers to $K^0$, $D^0$, $B^0$, and $B_s^0$. We focus on the contribution of the new $\Delta F=2$ FCNC operators to the mass difference, and study the mass bounds of the $SO(3)_f$ family gauge bosons.

This paper is organized as follows. In Section \ref{section_ffbar}, we review the mass difference of the neutral mesons in SM. In Section \ref{section_so3}, we introduce the effective Hamiltonian of $SO(3)$ family gauge interaction and list the corresponding hadronic matrix elements.  We give the numerical results in Section \ref{section_num} and a short conclusion in Section {\ref{section_conclusion}}.

\section{$F-\bar{F}$ mass difference in SM}\label{section_ffbar}

In this section, we give a brief review about the mass difference of $F-\bar{F}$ system. In SM if only the strong and electromagnetic interactions exist, i.e. $\mathcal{H}_0=\mathcal{H}_{S}+\mathcal{H}_{EM}$, $F$ and $\bar{F}$ would be stable as the eigenstates of the Hamiltonian $\mathcal{H}_0$ with same mass $m_0$. When the Hamiltonian $\mathcal{H}_W$ of weak interaction is considered, $F$ and $\bar{F}$ will mix together and decay to the same final states. $F-\bar{F}$ mixing is responsible for the mass difference between the mass eigenstates. In SM, it is known that the neutral meson mixing arises from the box diagrams through two $W$ boson exchange. The FCNC processes involve heavy quarks via loops and consequently they are perfect testing ground for heavy flavor physics. The small mass difference of the neutral $K$ and $B$ imposes severe constraints to new physics beyond SM, especially to those with FCNC at tree level.

The phenomenon of $F-\bar{F}$ mixing is important for it is relating to the $CP$ violation. With Wigner-Weisskopf
approximation\cite{wwa1,wwa2}, the wave function describing the oscillation and decay of $F-\bar{F}$ is
\begin{equation}
  |\psi(t)\rangle=\psi_1(t)|F\rangle+\psi_2(t)|\bar{F}\rangle,
\end{equation}
which evolves according to a Schr\"odinger-like equation
\begin{equation}
  i\frac{d}{dt}\left(\begin{array}{c}\psi_1(t)\\ \psi_2(t)\end{array}\right)=\mathcal{H}
  \left(\begin{array}{c}\psi_1(t)\\ \psi_2(t)\end{array}\right)\equiv\left(\mathcal{M}-\frac{i}{2}\Gamma\right)
  \left(\begin{array}{c}\psi_1(t)\\ \psi_2(t)\end{array}\right),
\end{equation}
where
\begin{eqnarray}
  \mathcal{M}&=&(\mathcal{H}+\mathcal{H}^\dag)/2,\\
  \Gamma&=&i(\mathcal{H}-\mathcal{H}^\dag).
\end{eqnarray}
The matrices $\mathcal{M}$ and $\Gamma$ are given, in second-order perturbation theory, by summing over intermediate
states $|n\rangle$,
\begin{eqnarray}
  &&\mathcal{M}_{ij}=m_0\delta_{ij}+\langle i|\mathcal{H}_W|j\rangle+\sum_n P\left\{
  \frac{\langle i|\mathcal{H}_W|n\rangle\langle n|\mathcal{H}_W|j\rangle}{m_0-E_n}\right\},\label{eq_hamiltonian}\\
  &&\Gamma_{ij}=2\pi\sum_n\delta(m_0-E_n)\langle i|\mathcal{H}_W|n\rangle\langle n|\mathcal{H}_W|j\rangle.
\end{eqnarray}
The result of the mixing is
\begin{equation}
  \Delta M-i\Delta\Gamma/2=2\sqrt{\mathcal{H}_{12}
  \mathcal{H}_{21}},
\end{equation}
where the $\Delta M$ ($\Delta \Gamma$) is the mass (width) difference.
The processes for $F-\bar{F}$ mixing in SM have been studied extensively and can be found in
\cite{bbarmixing1,bbarmixing2,bbarmixing3,bbarmixing4,bbarmixing5,bbarmixing6,bbarmixing7,kkbarmixing1,kkbarmixing2,ddbarmixing1,ddbarmixing2,ddbarmixing3}. Experimentally, the mass differences are\cite{PDG,ddexp1,ddexp2,ddbarlr}
 \begin{eqnarray}
\Delta M_{K}&=&(3.483\pm0.006) \times 10^{-15}\rm{GeV},\label{eq_kk_exp}\\
\Delta M_{B}&=&(3.337\pm0.033)\times 10^{-13}\rm{GeV},\label{eq_bb_exp}\\
\Delta M_{B_s}&=&(1.170\pm0.008)\times 10^{-11}\rm{GeV},\label{eq_bsbs_exp}\\
\Delta M_{D}&=&(1.4\pm0.5)\times10^{-14} \rm{GeV}.\label{eq_dd_exp}
\end{eqnarray}
The low energy effective interaction in SM provides a good approximation to $\Delta M_{B}$ and $\Delta M_{B_s}$, and also gives acceptable result of $\Delta M_K$ where the non-perturbation effect can not be ignored. However the value of $\Delta M_{D}$ in SM is only $10^{-18}\sim 10^{-17}$GeV. It is because the $D-\bar{D}$ system is different from the $B_{d,s}-\bar{B}_{d,s}$ and $K-\bar{K}$, where the internal quarks in the box diagrams are down-type quarks. Since the mass differences of the down-type quarks are much smaller than that of the
up-type quarks, the GIM mechanism\cite{GIM} works much efficiently for $D$ meson than that for $K$ and $B_{d,s}$ mesons. Furthermore, owing to the small Cabibbo-Kobayashi-Maskawa (CKM) angles the coupling to the third family is negligible, such that effectively only the
first two generations play a role, making the GIM mechanism even more efficient to $D-\bar{D}$ system.

As the short-distance effect in SM is small, it is suitable to study other effect in $D-\bar{D}$ system, such as the long-distance effects, which are difficult to calculate due to the non-perturbation, or some new physics effects. In this paper we assume the short-distance effect of new physics is dominated to $\Delta M_D$, and discuss the possible family gauge boson contribution.

\section{The effective Hamiltonian in $SO(3)_f$\label{section_so3}}

The Lagrangian about the interactions of quarks and the gauge bosons of $SO(3)_f$ is
\begin{eqnarray}
  \mathcal{L}_{SO(3)}=\frac{g_f}{2}\bar{\Psi}^U \mathcal{A}\!\!\!/ \Psi^U+\frac{g_f}{2}\bar{\Psi}^D
  \mathcal{A}\!\!\!/ \Psi^D,\label{eq_so3lagarangian}
\end{eqnarray}
where $g_f$ is the new gauge coupling and
\begin{eqnarray}
&&\Psi^U=\left(
\begin{array}{c}
u\\
c\\
t
\end{array}
\right), \quad \quad\quad \Psi^D=\left(
\begin{array}{c}
d\\
s\\
b
\end{array}
\right),\\
  &&\mathcal{A}_\mu= T^a A^a_\mu=i\left(\begin{array}{ccc}
0&-A^3&A^2\\
A^3&0&-A^1\\
-A^2&A^1&0
\end{array}
\right).
\end{eqnarray}
The $T^a(a=1,2,3)$ are the generators of the $SO(3)$ group and $A^a(a=1,2,3)$ are the gauge fields. After the
gauge symmetry breaking, the fermions get mass. To diagonalize the mass matrix of quarks, one needs transformations of the quark fields from interaction eigenstates to mass eigenstates,
\begin{equation}
  \Psi^{U,D}\to V_{U,D}\Psi^{U,D}.
\end{equation}
With the mass eigenstates, the Lagrangian can be re-expressed as
\begin{eqnarray}
  \mathcal{L}_{SO(3)}&=&\frac{g_f}{2}\bar{\Psi}^D V_D^\dag T^aA\!\!\!/^a V_D\Psi^D+
  \frac{g_f}{2}\bar{\Psi}^U V_U^\dag T^aA\!\!\!/^a V_U\Psi^U\nonumber\\
&\equiv&\frac{g_f}{2}\bar{\Psi}^D\left(V^1\sls{A}^1+V^2\sls{A}^2+V^3\sls{A}^3\right)\Psi^D+
\frac{g_f}{2}\bar{\Psi}^U\left(U^1\sls{A}^1+U^2\sls{A}^2+U^3\sls{A}^3\right)\Psi^U,\label{eq_lagarangian}
\end{eqnarray}
where
\begin{eqnarray}
V^c_{ij}=  -i\epsilon^{abc}(V^\dag_D)_{ia} (V_D)_{bj},\label{eq_down_mixing}\\
U^c_{ij}=  -i\epsilon^{abc}(V^\dag_U)_{ia} (V_U)_{bj}.\label{eq_up_mixing}
\end{eqnarray}
The $V^i$s and $U^i$s are not independent, since the $V_U$ and $V_D$ are combined to CKM matrix,
\begin{equation}
  V_{\rm{CKM}}=V_U^\dag V_D.\label{eq_ckm}
\end{equation}
In general, each of the three unitary matrixes $V_U$, $V_D$ or $V_{\rm{CKM}}$ has eight parameters, i.e., three mixing angles and five phases.
Only two unitary matrixes are independent according Eq.(\ref{eq_ckm}), thus there are 16 independent parameters to be input. In
SM, only the CKM matrix appears in the Lagrangian. Due to the re-phase invariance of the CKM matrix, only one phase has
physical effect, and the others can be moved to $V_U$ and $V_D$ which do not appear in SM. However, $V_U$ and $V_D$  appears in the new gauge interactions in Eq.(\ref{eq_so3lagarangian}).

From the Lagrangian given in Eq.(\ref{eq_lagarangian}), one can get
the effective Hamiltonian for the $\Delta F=2$ processes,
\begin{eqnarray}
  \mathcal{H}_{\rm{eff}}(\Delta F=2)&=&\mathcal{H}_{\rm{eff}}^{\rm{SM}}(\Delta F=2)+\mathcal{H}_{\rm{eff}}^{\rm{NP}}
  (\Delta F=2).
\end{eqnarray}
The $\mathcal{H}^{\rm {SM}}_{\rm eff}$ is from the
box diagrams in SM, while the $\mathcal{H}^{\rm{NP}}_{\rm{eff}}$ is
from the new gauge interactions,
\begin{eqnarray}
  \mathcal{H}_{\rm{eff}}^{\rm{NP}}(\Delta F=2)&=&C_1^{D-\bar{D}}(\bar{u} c)_V (\bar{u} c)_V+C_2^{D-\bar{D}}
  (\bar{u}_\alpha c_\beta)_V(\bar{u}_\beta c_\alpha)_V\nonumber\\
&&+C_1^{K-\bar{K}}(\bar{d} s)_V (\bar{d} s)_V+C_2^{K-\bar{K}}(\bar{d}_\alpha s_\beta)_V(\bar{d}_\beta s_\alpha)_V\nonumber\\
&&+C_1^{B-\bar{B}}(\bar{d} b)_V (\bar{d} b)_V+C_2^{B-\bar{B}}(\bar{d}_\alpha b_\beta)_V(\bar{d}_\beta b_\alpha)_V\nonumber\\
&&+C_1^{B_s-\bar{B}_s}(\bar{s} b)_V (\bar{s}
b)_V+C_2^{B_s-\bar{B}_s}(\bar{s}_\alpha b_\beta)_V(\bar{s}_\beta
b_\alpha)_V.\label{eq_ff_NP}
\end{eqnarray}
At tree level, the Wilson coefficients are
\begin{eqnarray}
 C_1^{D-\bar{D}}=g^2\left(\sum_{i=1}^{3}\frac{{U_{12}^i}^2}{4M_i^2}\right),
&&  C_1^{K-\bar{K}}=g^2\left(\sum_{i=1}^{3}\frac{{V_{12}^i}^2}{4M_i^2}\right),\nonumber\\
C_1^{B-\bar{B}}=g^2\left(\sum_{i=1}^{3}\frac{{V_{13}^i}^2}{4M_i^2}\right),
&&
C_1^{B_s-\bar{B}_s}=g^2\left(\sum_{i=1}^{3}\frac{{V_{23}^i}^2}{4M_i^2}\right).
\end{eqnarray}
Due to the color suppression, one has $C_2^{K-\bar{K}}=C_2^{B-\bar{B}}=C_2^{B_s-\bar{B}_s}=C_2^{D-\bar{D}}=0$ at tree level.

The $\mathcal{M}_{12}^{\rm{NP}}$ is the contribution from the new
operators whose hadronic matrix elements are given by,
\begin{eqnarray}
  \langle VV\rangle&=&\frac{1}{4}\langle\left[(V-A)+(V+A)\right] \left[(V-A)+(V+A)\right]\rangle,\nonumber\\
&=&\frac{1}{2}\left[\langle(V-A)(V-A)\rangle+\langle(V-A)(V+A)\rangle\right].\label{eq_vv_factorization}
\end{eqnarray}
The factorizations of the hadron matrix elements are listed as follows,
\begin{eqnarray}
 \langle\bar{F}|(V-A)(V-A)|F\rangle&=&\frac{4}{3}m_Ff_F^2B^F_1(\mu),\\
  \langle\bar{F}|(V-A)(V+A)|F\rangle&=&-\frac{2}{3}R(\mu)m_Ff_F^2B^F_2(\mu),
\end{eqnarray}
where the factor $R={M}^2/{(m_q+m_q^\prime)^2}$  and the $M$ is the average mass of the $F$ and $\bar{F}$. $m_q$ and $m_{q^\prime}$ are the mass of the quarks which are the components of the meson. $f_F$ is the decay constant and $B_i$s are the bag parameters which are unit in naive factorization\cite{factorization1,factorization2}.
%
It should be noted that to get Eq.(\ref{eq_vv_factorization}) the following equations have been used,
\begin{eqnarray}
  &&\langle(V-A)(V-A)\rangle=\langle(V+A)(V+A)\rangle,\\
  &&\langle(S-P)(S-P)\rangle=\langle(S+P)(S+P)\rangle.
\end{eqnarray}
 As one knows the strong interaction preserves the chiral symmetry, the hadronic matrix elements of right-hand
 operators will be the same as the left-hand operators.

\begin{table}[htb]
\tbl{The average mass, the decay constant  and the bag parameters of the neutral mesons.}
{\begin{tabular}{c|c c c c}
\toprule
 & K & D & B &$B_s$\\
\hline
$M_F$ &0.498GeV & 1.86 GeV &5.28GeV & 5.37GeV\\
\hline
$f_F$ &   $156\pm0.8$MeV& $191\pm23$MeV& $220\pm40$MeV  & $205\pm10$MeV  \\
\hline
 $B_1$ & $0.571\pm0.048$& $0.87 \pm 0.03$ & $0.87\pm 0.04$& $0.86\pm 0.02$\\
 \hline
 $B_2$& $0.562\pm0.039$& $1.46 \pm 0.09$ & $1.91\pm 0.04$ & $1.94\pm0.03$\\
 \hline
 \end{tabular}\label{table_input}}
\end{table}
\section{Numerical results\label{section_num}}
In our calculation, the following inputs are adopted.
$m_t=171.2$GeV, $M_W=80.4$GeV, $\alpha_s(M_Z)=0.118$, and the
mass\cite{PDG}, decay constants\cite{fb,fd} and the bag parameters \cite{Babich:2006bh,d_bag1,d_bag2,bag1,bag2,Becirevic:2001xt}
are listed in
Table.(\ref{table_input}). The CKM matrix $V_{\rm CKM}$ is
parameterized\cite{wolfenstain} with $A=0.806$, $\lambda=0.2272$, $\bar{\rho}=0.195$,
$\bar{\eta}=0.326$. With such inputs, the results of the mass differences $\Delta M_F$ in SM are listed as follows,
 \begin{eqnarray}
   &&\Delta M_K=2.312^{+0.024+0.466}_{-0.024-0.462}\times10^{-15}{\rm GeV},\label{eq_kk_sm}\\
   &&\Delta M_{B_d}=3.483^{+0.991+0.161}_{-0.789-0.159}\times10^{-13}{\rm GeV},\label{eq_bb_sm}\\
   &&\Delta M_{B_s}=1.20^{+0.47+0.03}_{-0.77-0.03}\times10^{-11}{\rm
   GeV}.\label{eq_bsbs_sm}
 \end{eqnarray}
The first uncertain comes from the decay constant, and the second comes from the bag parameter. One can find that
the results are close to the experimental values. In the following we consider the new physics effect in Eq.(\ref{eq_ff_NP}).

\subsection{The results with $V_{D}\sim1$}
\begin{figure}[htb]
\centerline{\psfig{file=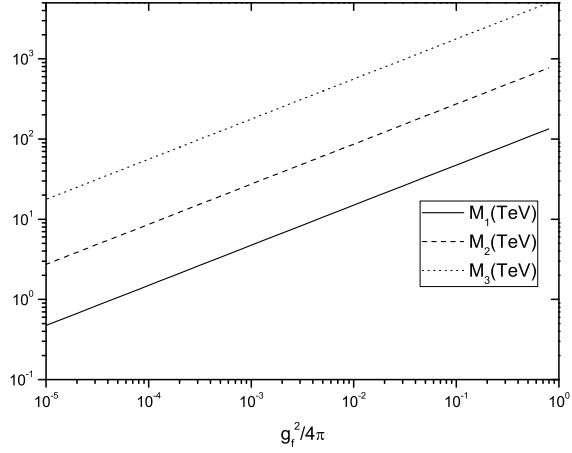,width=10.0cm}}
\caption{The low boundary of the new gauge bosons' mass as functions of the new coupling constant $\frac{g_f^2}{4\pi}$. \label{fig_bound}}
\end{figure}

First, for simplicity one can assume the mixing matrix of the down-type quarks is unit, and the the mixing matrix of up-type quarks is
$V_U=V^\dag_{\rm{CKM}}$. The free parameters are only the gauge coupling $g_f$ and the mass of the gauge bosons. Since the
effect of new physics is proportional to the $g_f^2/M^2_i$, one can get the upper limits from $\Delta M_{B_s,B_d,K}$,
\begin{eqnarray}
  \frac{g_f^2}{M_1^2}\leq1.4\times 10^{-10}{\rm GeV}^{-2},\nonumber\\
  \frac{g_f^2}{M_2^2}\leq4.2\times 10^{-12}{\rm GeV}^{-2},\label{eq_bound}\\
  \frac{g_f^2}{M_3^2}\leq1.0\times 10^{-13}{\rm GeV}^{-2}.\nonumber
\end{eqnarray}
If the gauge interaction coupling constant is fixed, the above limits give lower limits of the gauge boson mass as plotted in Fig.(\ref{fig_bound}).
When the new gauge coupling constant is taken as strong as the weak
interaction in SM, $g_f=g=0.657$, one can get the mass limits
which are $M_1\geq27$TeV, $M_2\geq160$TeV, and $M_3\geq 10^3$TeV.
So we get the scale of the new interactions about
$\mu_f\sim 10^2$TeV, in addition to $\Lambda_{\rm QCD}\sim 10^2$MeV and $\mu_{\rm EW}\sim 10^2$GeV in SM.
We take the limits in Eq.(\ref{eq_bound}) as input and get the
$D-\bar{D}$ mass difference
\begin{equation}
  \Delta M_D=0.85^{+0.08+0.04+0.12}_{-0.08-0.03-0.11}\times
  10^{-14}{\rm GeV},
\end{equation}
where the first uncertain is from the decay constant and the others are from the bag parameters $B_1$ and $B_2$. One can find that our result is consistent with the experimental values.

\subsection{$D-\bar{D}$ and $K-\bar{K}$}
Finally, we consider another approximation. As the off-diagonal elements of CKM matrix are small,
\begin{equation}
V_{\mbox{CKM}}={1}+\mathcal{O}(\lambda),
\end{equation}
the $V_U$ and $V_D$ are equal at leading order,
\begin{equation}
  V_U=V_D+\mathcal{O}(\lambda).
\end{equation}
According to Eq.(\ref{eq_down_mixing},\ref{eq_up_mixing}), one can get
\begin{equation}
  U^i=V^i+\mathcal{O}(\lambda), \ (i=1,2,3).
\end{equation}
The $D-\bar{D}$ is related with $U^i_{12}$ which is expressed as
\begin{equation}
  U_{12}^i=V_{12}^i+\mathcal{O}(\lambda)
\end{equation}
and the $K-\bar{K}$ is related with $V^i_{12}$.

\begin{figure}[htb]
\begin{center}
  \includegraphics[scale=0.25]{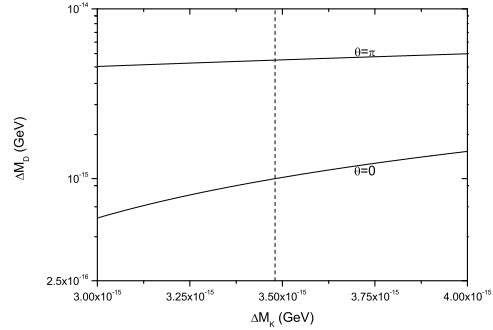}
  \caption{The $\Delta M_{D}$ and $\Delta M_{K}$. The dashed line is the experimental value of $\Delta M_K$}\label{fig_ddkk}
  \end{center}
\end{figure}

 At the leading order, as $U^i=V^i$ one has Wilson coefficients
\begin{eqnarray}
&&  C_2^{D-\bar{D}}=C_2^{K-\bar{K}}=0,\\
&&  C_1^{D-\bar{D}}\approx
C_1^{K-\bar{K}}=g^2\left(\sum_{i=1}^{3}\frac{{V_{12}^i}^2}{M_i^2}\right)\equiv
C e^{i\theta}.
\end{eqnarray}
The imaginary part of the coefficient is relating with the $CP$ violation in
$K-\bar{K}$ system which is detected\cite{PDG} as $|\epsilon_K|=
(2.229\pm0.012)\times10^{-3}$. The imaginary part is small which implies the phase $\theta$ is close to $0$
or $\pi$. In Fig.(\ref{fig_ddkk}), the plot of the $\Delta M_{D}$ and $\Delta M_{K}$ with the same $C$ is shown. From the plot one can get
two solutions,
\begin{eqnarray}
  \theta=0,&& C\sim1.4\times 10^{-14}{\rm GeV}^{-2},\\
  \theta=\pi,&&C\sim6.7\times 10^{-14}{\rm GeV}^{-2}.
\end{eqnarray}
Taking the two solutions as input, one can get the $\Delta M_{D}$,
\begin{eqnarray}
 \theta=0,&& \Delta M_{D}=0.1\times 10^{-14}GeV,\nonumber\\
  \theta=\pi,&&\Delta M_{D}=0.5\times 10^{-14}GeV.
\end{eqnarray}
\section{Conclusion\label{section_conclusion}}
As one knows that if the mass of the SM Higgs lies between 130 and 200GeV\cite{higgsarg}, the SM can be valid at energy scales all the
way up to Plank scale. But it seems  unnatural if there is a very large desert from Electro-Weak scale to Plank scale. When the family gauge symmetry is introduced, there are FCNC at tree level especially the $\Delta F=2$ processes and more CP violation sources which are very interesting to flavor physics. In this paper we study the effect of the new gauge interaction to the mass difference of the $F-\bar{F}$ systems. We get the scale of the new gauge interaction at about $10^2$TeV in addition to the the gauge scales in SM. With the constrain from $K-\bar{K}$ and $B-\bar{B}$, a result of $\Delta M_D$ consistent to the experimental data is gotten. In this paper, only the new physics effect at tree level is considered to get a simple constrain to the scale of the new gauge bosons. To get more precise results, the QCD corrections to the new operators and  the long-distance effect should be considered.

\section*{Acknowledgments}
The authors would like to thank Prof. Yue-Liang Wu for useful discussions and suggestions. This work was
supported in part by the National Science Foundation
of China, China Postdoctoral Science Foundation (CPSF), Natural Science Foundation of Shandong
Province(\#ZR2011AQ013,\#JQ200902) and Independent Innovation Foundation of Shandong University.

\end{document}